\documentclass{article}
\usepackage{filecontents,catchfile}
\usepackage{spconf,amsmath,graphicx}

\usepackage{dsfont}
\usepackage{amsthm,amssymb}
\usepackage{amsmath,etoolbox}
\usepackage{mathtools}

\usepackage{enumitem}
\setlist{nosep, leftmargin=14pt}

\usepackage{mwe} % to get dummy images

\title{Few-shot image segmentation for cross-institution male pelvic organs using registration-assisted prototypical learning}
\name{
\begin{tabular}{@{}c@{}}
Yiwen Li$^{1}$ \qquad
Yunguan Fu$^{2,3}$ \qquad
Qianye Yang$^{2}$ \qquad
Zhe Min$^{2}$ \qquad
Wen Yan$^{2,4}$\\
Henkjan Huisman$^{5}$ \qquad
Dean Barratt$^{2}$ \qquad
Victor Adrian Prisacariu$^{1}$ \qquad
Yipeng Hu$^{1,2}$ \qquad
\end{tabular}
}

\address{$^{1}$ University of Oxford \qquad
$^{2}$ University College London \qquad
$^{3}$ InstaDeep\\
$^{4}$ City University of Hong Kong \qquad
$^{5}$ Radboud University Nijmegen Medical Centre}

\begin{document}
%\ninept
%

\maketitle
\begin{abstract}
The ability to adapt medical image segmentation networks for a novel class such as an unseen anatomical or pathological structure, when only a few labelled examples of this class are available from local healthcare providers, is sought-after. This potentially addresses two widely recognised limitations in deploying modern deep learning models to clinical practice, expertise-and-labour-intensive labelling and cross-institution generalisation. This work presents the first 3D few-shot interclass segmentation network for medical images, using a labelled multi-institution dataset from prostate cancer patients with eight regions of interest. We propose an image alignment module registering the predicted segmentation of both query and support data, in a standard prototypical learning algorithm, to a reference atlas space. The built-in registration mechanism can effectively utilise the prior knowledge of consistent anatomy between subjects, regardless whether they are from the same institution or not. Experimental results demonstrated that the proposed registration-assisted prototypical learning significantly improved segmentation accuracy (p-values$<$0.01) on \textit{query} data from a holdout institution, with varying availability of \textit{support} data from multiple institutions. We also report the additional benefits of the proposed 3D networks with 75\% fewer parameters and an arguably simpler implementation, compared with existing 2D few-shot approaches that segment 2D slices of volumetric medical images.

\end{abstract}
\section{Introduction}
\label{sec:intro}
Magnetic resonance (MR) imaging is of growing importance in the clinical pathway for diagnosing and treating prostate cancer~\cite{scheidler1999prostate,dickinson2013image}, with which, MR has been proposed for planning a number of surgical and interventional procedures, such as targeted biopsy~\cite{hamid2019smarttarget}, focal ablation, radiation therapy~\cite{lee2003radiotherapy} and robotic surgery~\cite{mcclure2012use}. In addition to detecting cancerous lesions on multi-parametric MR images, delineating the prostate gland, its zonal structures, and surrounding vulnerable and/or functional regions of interest (ROIs) is also required, and is feasible on T2-weighted scans, for planning these procedures. Labelling these anatomical structures for training a fully-supervised segmentation network requires expert knowledge of anatomy and uroradiology. Furthermore, the lack of generalisation ability of modern image segmentation networks~\cite{gibson2018inter} complicates their translation into clinical practice, with curating training datasets for individual institutions seemly neither feasible nor cost-effective. 

Considering the multi-institution context, this work investigates a number of practical scenarios, in which the types of ROIs that require segmentation at a local institution may or may not be different from those in the network training dataset. This could be due to the availability of labelled scans, required ROIs for different types of procedure or institution-specific imaging protocols. 

Segmenting multiple structures from MR images of the male pelvis is a typical example of multi-class segmentation applications, in which few-shot learning can help generalise to unseen classes (here, anatomical or pathological structures) and institutions with only a few annotations. Specifically, the model should be capable of segmenting a \emph{novel class} from a \emph{query} image acquired from a \textit{novel institution}, given only a \emph{support} image and its binary mask of the novel class from an institution where the labelled data are available.

In this work, we adopt the reference-quality few-shot episodic training to acquire inter-class generalisation, described in Secs.~\ref{sec:episodic} and~\ref{sec:prototype}. For inter-institution generalisation, we propose a 3D neural network embedded with an image alignment module to align the \emph{support} and \emph{query} images prior to the segmentation module, motivated by medical-image-specific observations of the difference between intra- and inter-institution data characteristics, due to different scanners and local imaging protocols, discussed further in Sec.~\ref{sec:align}. Our contributions are summarised as follows.

\begin{itemize}
    \item We present a new multi-class segmentation application for planning prostate cancer procedures, using a labelled MR dataset with 8 ROIs from 178 patients across 6 institutions, and report a baseline supervised performance.
    \item We propose the first 3D neural network for prototypical few-shot multi-class segmentation, which requires fewer parameters while achieving state-of-the-art performance.
    \item We describe a novel medical-image-specific alignment module and demonstrate its efficacy in improving generalisation across data from different institutions. 
\end{itemize}

\section{Related Work}
\label{sec:related_work}
The task of few-shot segmentation was first discussed in vision applications~\cite{shaban2017one} and has since been adapted into the medical imaging field~\cite{roy2020squeeze}.
The commonly adopted approach is prototypical learning~\cite{dong2018few}, where the foreground features of the \emph{support} image is pooled into a \textit{prototype} vector to represent the novel class. The prototype is subsequently compared with the \emph{query} images that require segmentation. Although the pooling enables reducing computation, it may lead to a loss of local intraclass information. Both ALPNet~\cite{ouyang2020self} and LSNet~\cite{yu2021location} achieved improvement by extracting multiple local-prototypes from the support image, with the assumption that support and query images share a similar spatial layout. 
To the best of our knowledge, previously proposed prototypical few-shot segmentation methods for medical images are based on 2D networks~\cite{roy2020squeeze,ouyang2020self,yu2021location}. 
Other non-few-shot approaches include segmenting multiple organs for planning radiotherapy on CT~\cite{lei2020male}, and segmenting the prostate gland and its zonal structures on MR~\cite{rundo2019use,litjens2014evaluation,liu2019automatic,roth2021federated}.

\section{Cross-Institution Few-Shot Task}
\label{sec:task}
% symbols
% image I
% mask M
% support s
% query q
% institution u
In this section, we define the multi-class multi-institution segmentation task of interest in this work. Let $(I, M(c), u)$ be an image-mask-institution trio, where $M(c)$ represents the binary mask of a class $c\in\mathcal{C}$ in the image $I$ acquired from an institution $u\in\mathcal{U}$. The classes $\mathcal{C}$ and institutions $\mathcal{U}$ are divided into disjoint sets $(\mathcal{C}_\text{base}, \mathcal{C}_\text{novel})$ and $(\mathcal{U}_\text{base}, \mathcal{U}_\text{novel})$, respectively. Now let \emph{base dataset} and \emph{novel dataset} be respective $\mathcal{D}_\text{base} = \{(I_i, M_i(c), u)|c\in\mathcal{C}_\text{base}, u\in\mathcal{U}_\text{base}\}^{N_\text{base}}_{i=1}$ and $\mathcal{D}_\text{novel} = \{(I_j, M_j(c))|c\in\mathcal{C}_\text{novel},u\in\mathcal{U}\}^{N_\text{novel}}_{j=1}$, such that no class is labelled in both datasets and novel institutions do not occur in the base dataset. The task is to train a model on the base dataset, such that it learns to segment a novel class with only a few support examples. 
Specifically, following \cite{roy2020squeeze, ouyang2020self}, the evaluation is performed in an episodic one-shot manner: each \emph{episode} samples a novel class $c \in C_\text{novel}$ and a \emph{support} $(I^s, M^s(c), u^s)$ and \emph{query} $(I^q, M^q(c), u^q)$ pair, where $u^s$ and $u^q$ can be the same or different institutions. In this work, we mainly focus on the case $u^q\in\mathcal{U}_\text{novel}$, i.e. the query image comes from the novel institution. The model is trained to predict the query mask $\hat{M}^q(c)$ for a query image $I_q$, given a support image-mask pair $(I^s, M^s(c))$.

\section{Data}\label{sec:data}
A total of 178 T2-weighted MR images, from 6 institutions, were acquired from the same number of prostate cancer patients. The cross-institution imaging protocols contain multiple scanners (manufacturers and 1.5/3T), varying field-of-view and anisotropic voxels, in-plane voxel dimensions ranging between 0.3 and 1.0 mm and out-of-plane spacing between 1.8 and 5.4 mm. Eight anatomical structures of planning interest were identified and manually segmented by five biomedical imaging researchers, each annotating a mixed-institution subset using an institution-stratified sampling. These structures include the prostate transition zone, peripheral zone, seminal vesicles, neurovascular bundles, obturator internus muscle, rectum, bladder, and pelvic bones that are visible in the MR field-of-view. All images were resampled and centre-cropped to an image size of $256\times256\times48$, with a voxel dimension of $0.75\times0.75\times2.5$.

\section{Method}
\label{sec:method}

\begin{figure*}
    \centering
    \centerline{\includegraphics[width=\linewidth]{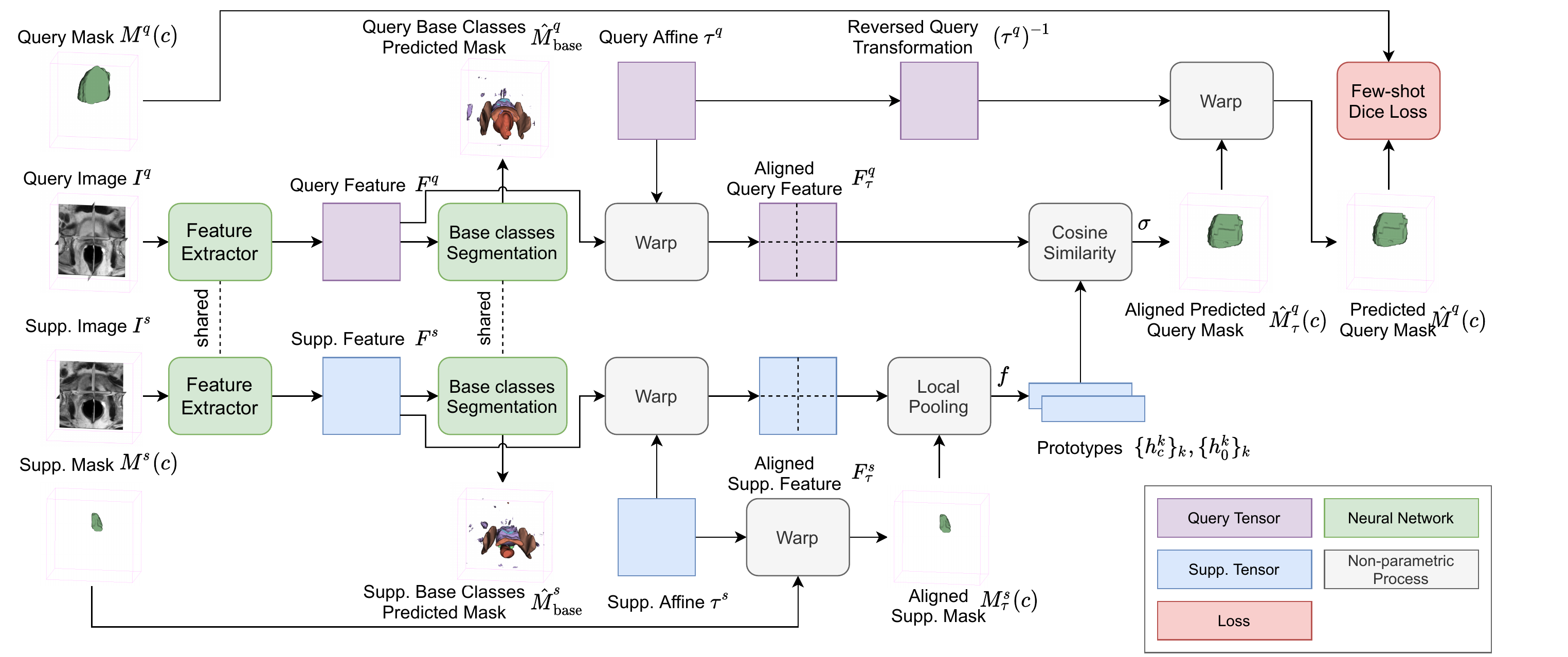}}
    \caption{Few-shot pipeline, where the query and support base classes are input to the image alignment module, from which the query and support transformations are outputted. Warp module transforms the image or mask with the affine transformation.}
    \label{fig:few_shot_pipeline}
\end{figure*}

\begin{figure}
    \centering
    \centerline{\includegraphics[width=\linewidth]{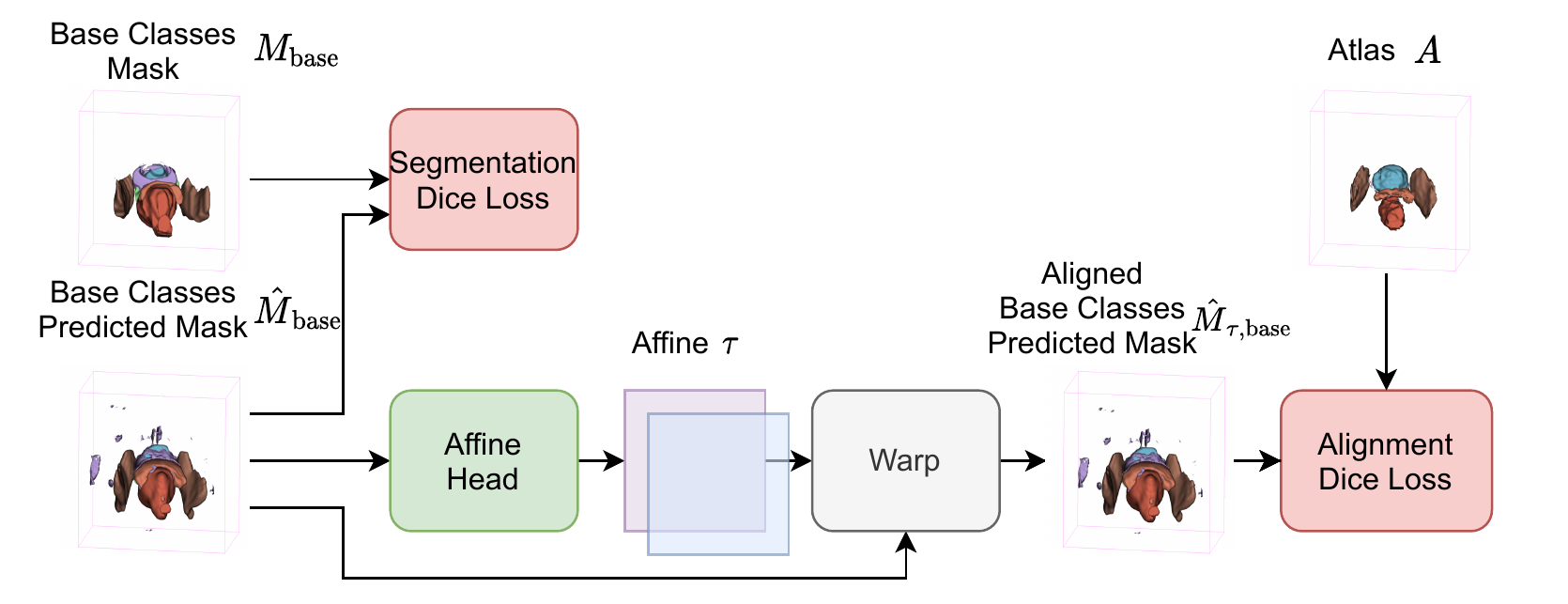}}
    \caption{Image alignment module.}
    \label{fig:alignment_pipeline}
\end{figure}

\subsection{Episodic Few-shot Training}
\label{sec:episodic}
% need to explain normal few shot data flows
% Like many state-of-the-art few-shot semantic segmentation methods, this work also 
This work adopts the episodic training paradigm. An overview of the proposed approach is illustrated in Fig. \ref{fig:few_shot_pipeline}, where an \emph{episode} consists of a support $(I^s, M^s(c), u^s)$ and query $(I^q, M^q(c), u^q)$ pair, for a base class $c$ sampled from $C_\text{base}$.

The query and support images $I^s$ and $I^q$ are encoded by a shared feature extractor into support and query feature maps, $F^s$ and $F^q$, of the same shape.
The class prototype $h_c$ and background prototype $h_0$ are then derived by averaging $F^s$ over voxels of the class $c$ (those being $1$) and background (those being $0$), respectively, i.e. $h_c = f(F^s, M^s(c))$ and $h_0 = f(F^s, 1-M^s(c))$, where
\begin{align}\label{eq:prototype}
    f(F, M) = \frac{\sum\limits_{(x,y,z)} F_{(x,y,z)} M_{(x,y,z)}}{\sum\limits_{(x,y,z)}{M_{(x,y,z)}}},
\end{align}
with $(x,y,z)$ iterating over all voxels.
Each voxel in the query mask $\hat{M}^q(c)$ is predicted using a softmax between the features $h_c$ and $h_0$, i.e. 
\begin{align}\label{eq:query-mask-pred}
    \hat{M}^q(c)_{(x,y,z)} = \sigma(F^q_{(x,y,z)}, h_c, h_0) = \frac{\exp(s_c)}{\exp(s_c)+\exp(s_0)},
\end{align}
where $s_\star=\frac{F^q_{(x,y,z)}\cdot h_\star}{\|F^q_{(x,y,z)}\| \|h_\star\|}, \star\in\{c,0\}$ is the cosine similarity between the feature vector $F^q_{(x,y,z)}$ and prototype $h_\star$. The Dice loss is used between the predicted query mask $\hat{M}^q(c)$ and ground-truth $M^q(c)$, denoted as $\mathcal{L}_\text{few-shot}$.

\subsection{Local Prototype}
\label{sec:prototype}
Following LSNet~\cite{yu2021location}, overlapping windows $\{G_k\}_{k=1}^K$ of size $\alpha_w W \times \alpha_h H \times \alpha_d D$ are sampled in 3D, partitioning images of spatial size $W\times H \times D$, with the equidistant spacing between window centres being half the window size. For each window $G_k$, two local prototype feature vectors $h_c^k$ and $h_0^k$ are calculated via Eq.~\eqref{eq:prototype} by iterating $(x,y,z)$ inside the window $G_k$.
For the predicted query mask, each feature vector is compared with all local prototype feature vectors using Eq.~\eqref{eq:query-mask-pred}, for $\star\in\{c,0\}$, now with the maximum $s_\star=\max\limits_{k}\frac{F^q_{\tau,(x,y,z)}\cdot h_\star^k}{\|F^q_{\tau,(x,y,z)}\| \|h_\star^k\|}$.

\subsection{Image Alignment Module}
\label{sec:align}
% here we add the alignment
As discussed in Sec.~\ref{sec:intro}, difference in intra- and inter-institution variations may challenge this few-shot segmentation task described in Sec.~\ref{sec:task}.
For example, most anatomical structures are located consistently, yet their absolute locations vary substantially between scans acquired from different institutions with varying image size, orientation, and voxel dimensions. 

This observation motivated this work to spatially align the query $I^q$ and support images $I^s$ to a reference space, represented by a mask atlas $A$, illustrated in Fig.~\ref{fig:alignment_pipeline}. The registration process is conjectured to alleviate the discrepancy of the data across institutions and therefore reduce the amount of cross-institution training data required for generalisation. In this work, we consider an affine transformation to account for the above-discussed spatial difference with potentially variable imaging calibration uncertainties, although higher-degree transformation will also be of interest. Furthermore, to avoid repeated feature map extraction, we propose to apply the transformation directly on feature maps ($F^q$ and $F^s$), rather than on images.
As in Figs.~\ref{fig:few_shot_pipeline}, a shared segmentation head segments all the base classes from the query and support feature maps, $F^q$ and $F^s$. Denote the multi-class prediction as $\hat{M}^q_\text{base}$ and $\hat{M}^s_\text{base}$. The corresponding ground-truth are multi-class masks $M^q_\text{base}$ and $M^s_\text{base}$. The segmentation head is trained with an additional Dice-based loss:
\begin{align}
    L_\text{seg}= L(\hat{M}^q_\text{base}, M^q_\text{base}) + L(\hat{M}^s_\text{base}, M^s_\text{base}),
\end{align}
where $L$ represents the Dice loss function.

Affine transformations $\tau^q$ and $\tau^s$ are predicted for $\hat{M}^q_\text{base}$ and $\hat{M}^s_\text{base}$, before being applied to both the feature maps and labels for query and support, respectively. Denoting the transformed feature maps and labels as $(F_\tau^q, M_\tau^q(c))$ and $(F_\tau^s, M_\tau^s(c))$, with
the respective prototypes being $h_{c,\tau} = f(F_\tau^s, M_\tau^s(c))$ and $h_{0,\tau} = f(F_\tau^s, 1-M_\tau^s(c))$, the query mask $\hat{M}^q_\tau(c)$ can be predicted with $F_\tau^q$, $h_{c,\tau}$ and $h_{0,\tau}$, as in Eq.~\eqref{eq:query-mask-pred}. The final query mask prediction $\hat{M}^q(c)=(\tau^q)^{-1}\circ\hat{M}^q(c)$ by applying the reversed affine transformation $(\tau^q)^{-1}$.

The proposed transformation head is trained with a loss between the aligned predicted mask $\hat{M}^*_{\tau,\text{base}}=\tau^*\circ \hat{M}^*_\text{base}$, $*\in\{q,s\}$ and a pre-defined mask atlas $A$ that segments all base classes for both query and support,
\begin{align}
    L_\text{align}= L(\hat{M}_{\tau,\text{base}}^q, A) + L(\hat{M}_{\tau,\text{base}}^s, A)
\end{align}
The details of the proposed image alignment module are also illustrated in Fig. \ref{fig:alignment_pipeline}.

\begin{figure*}[t]
    \centering
    \centerline{\includegraphics[width=0.8
    \linewidth]{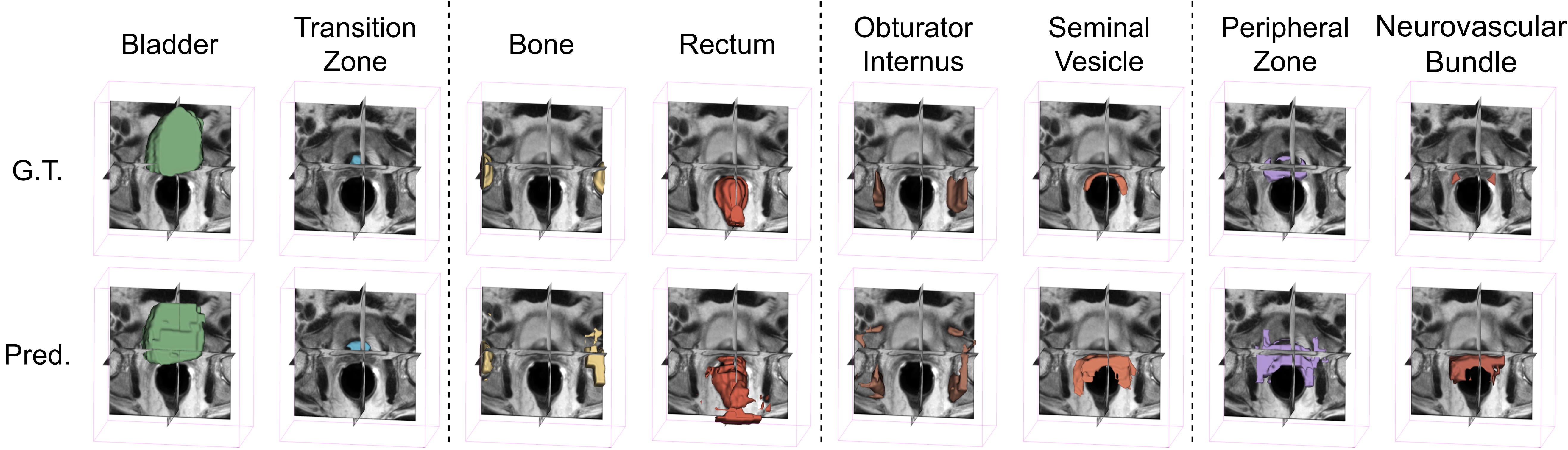}}
    \caption{Example holdout results comparing the ground-truth (G.T.) and the predicted (Pred.) segmentation.}
    \label{fig:qualitative}
\end{figure*}

\section{Experiment and Results}
%\subsection{Implementation Details}
Using the dataset described in Sec.~\ref{sec:data}. One out of 6 available institutions was used as the novel institution, which contains 21 subjects. For the remainder five base institutions, images from each institution were randomly divided into training and test sets with a 3:1 ratio. All were labelled with 8 classes, randomly divided into four folds: Folds 1 (bladder, transition zone), 2 (bone, rectum), 3 (obturator internus, seminal vesicle) and 4 (peripheral zone, neurovascular bundle).

In a given fold, these above-specified classes were considered as novel classes, the other three folds representing base classes. Base dataset was thus formed with training images from the five base institutions and masks from the base classes, while the novel dataset was formed by test images from the five base institutions, all images from the novel institution and masks from the novel classes.
For each of the fold-specified novel classes, all images from the novel institution were considered query images for evaluation. A binary Dice for this novel class was calculated between each query image with a sampled support image from each of the six institutions, excluding the query. The results were summarised when support images from 1) all institutions, 2) base institutions, and 3) novel institutions, denoted by `s\_ins'.

We report the results from a 2D baseline `2d' (LSNet~\cite{yu2021location}), the proposed `3d' (described in Sec.~\ref{sec:prototype} without the image alignment module), `3d\_seg' (with only the segmentation head but without the alignment head), and the proposed `3d\_seg\_align' (with the image alignment module). 
The 3D networks adopted a 3D UNet for feature extraction, and 3D windows with $\alpha_w$=$\alpha_h$=1/8, $\alpha_d$=1/4. The atlas was an average over masks from a randomly sampled base institution.
A fully supervised model was trained on the base dataset images with masks from both the base and novel classes. The results on the novel institution images are reported as an ``upper-bound'' performance.
Random rotation, translation and scaling were applied for data augmentation in all training.

\newcommand{\foldresult}{\small
\begin{tabular}{|cc|cccc|c|}
\hline
method          & s\_ins                & fold1     & fold2     & fold3     & fold4     & mean\\
\hline
                & all                   & 42.04     & 46.26     & 35.07     & 19.63     & 34.62\\        
2d              & base                  & 41.28     & 44.40     & 29.93     & 19.06     & 33.67\\
                & novel                 & 45.07     & 53.69     & 33.10     & 21.90     & 38.44\\
\hline
                & all                   & 41.97     & 46.23     & 31.00     & 19.64     & 34.71\\        
3d              & base                  & 41.26     & 44.55     & 30.26     & 19.23     & 33.83\\
                & novel                 & 44.79     & 52.92     & 33.97     & 21.29     & 38.24\\
\hline
                & all                   & 48.57     & 48.35     & 31.75     & 23.46     & 38.03\\        
3d\_seg         & base                  & 48.83     & 46.12     & 30.33     & 22.64     & 36.98\\
                & novel                 & 47.49     & 57.25     & 37.40     & 26.76     & 42.22\\
\hline
                & all                   & 47.80     & 49.81     & 33.30     & 26.10     & 39.25\\        
3d\_seg\_align  & base                  & 47.81     & 48.00     & 32.23     & 26.49     & 38.63\\
                & novel                 & 47.75     & 57.03     & 37.60     & 24.56     & 41.73\\
\hline
\multicolumn{2}{|c|}{fully supervised}  & 74.78     & 73.71     & 66.18     & 42.60     & 64.31\\
\hline
\end{tabular}}
\begin{table}[!ht]
    \centering
    \caption{Dice (\%) on query images from the novel institution.}
    \small
\begin{tabular}{|cc|cccc|c|}
\hline
method          & s\_ins                & fold1     & fold2     & fold3     & fold4     & mean\\
\hline
                & all                   & 42.04     & 46.26     & 35.07     & 19.63     & 34.62\\        
2d              & base                  & 41.28     & 44.40     & 29.93     & 19.06     & 33.67\\
                & novel                 & 45.07     & 53.69     & 33.10     & 21.90     & 38.44\\
\hline
                & all                   & 41.97     & 46.23     & 31.00     & 19.64     & 34.71\\        
3d              & base                  & 41.26     & 44.55     & 30.26     & 19.23     & 33.83\\
                & novel                 & 44.79     & 52.92     & 33.97     & 21.29     & 38.24\\
\hline
                & all                   & 48.57     & 48.35     & 31.75     & 23.46     & 38.03\\        
3d\_seg         & base                  & 48.83     & 46.12     & 30.33     & 22.64     & 36.98\\
                & novel                 & 47.49     & 57.25     & 37.40     & 26.76     & 42.22\\
\hline
                & all                   & 47.80     & 49.81     & 33.30     & 26.10     & 39.25\\        
3d\_seg\_align  & base                  & 47.81     & 48.00     & 32.23     & 26.49     & 38.63\\
                & novel                 & 47.75     & 57.03     & 37.60     & 24.56     & 41.73\\
\hline
\multicolumn{2}{|c|}{fully supervised}  & 74.78     & 73.71     & 66.18     & 42.60     & 64.31\\
\hline
\end{tabular}
    \label{tab:novel_result}
\end{table}

%results
As summarised in Table~\ref{tab:novel_result} and illustrated in Fig.~\ref{fig:qualitative}, the segmentation accuracy from the proposed `3d\_seg\_align' has seen significant improvements (p-value$<$0.01) over the previously proposed `2d' baseline, consistent across the tested scenarios. 
When support images are from all institutions, the `3d' network achieved a Dice score of $34.71\%$, similar to the 2D baseline. With the proposed image alignment module, adding the segmentation head and the affine head subsequently led to absolute improvements of $3.32\%$ (p-value=0.009) and $1.22\%$ (p-value=0.013).
All 3D networks reduced the number of parameters from $23.5$ million (the `2d' baseline) to $5.7$ million - a $\sim$75\% reduction. Interestingly, the 2D baseline and the proposed 3D network without image alignment module saw decreased performance by at least $12\%$, when the query and support images come from different institutions, whilst this gap was substantially reduced with the proposed image alignment module in many tested cases.

\section{Conclusion}

The above-reported cross-institution generalisation improvement, reduced network size, and decreased sensitivity to support the data source all bring considerable clinical benefits to planning the prostate cancer procedures discussed in Sec.~\ref{sec:intro}. The remaining performance gap between few-shot algorithms and the fully supervised remains an open research challenge and warrants future methodological or translational studies, such as non-affine alignment or data-efficient training. 

\section{Compliance with Ethical Standards}

This research study was conducted retrospectively using human subject data made available in open access by The Cancer Imaging Archive and Prostate MR Image Database and the PROMISE12 Challenge. UCL study data were acquired inline with ethics approvals granted by London-Dulwich and UCLH-NHS Research Ethics Committees (14/LO/0830, 19/LO/1129).

\section{Acknowledgement}

This work is supported by the Wellcome/EPSRC Centre for Interventional \& Surgical Sciences [203145Z/16/Z] and the CRUK International Alliance for Cancer Early Detection (ACED) [C28070/A30912; C73666/A31378].

\bibliographystyle{IEEEbib}
\bibliography{ref}

\end{document}